# Investigating the dispersion state of alumina suspensions: contribution of Cryo-FEGSEM characterizations

Audrey Lasalle[*], Christian Guizard, Sylvain Deville

Laboratoire de Synthèse et Fonctionnalisations des Céramiques,
UMR 3080 CNRS/Saint-Gobain, 84306 Cavaillon, France

Fabrice Rossignol

Laboratoire de Science des Procédés Céramiques et de Traitements de Surface,
UMR CNRS 6638, ENSCI, 87065 Limoges cedex, France

Pierre Carles

Service de Microscopie électronique, Université de Limoges, 87065 Limoges Cedex, France

**Abstract**

We illustrate in this paper the interest of cryo scanning electron microscopy (cryo-FEGSEM) to investigate the stability of alumina suspensions. The stability is investigated through viscosity, zeta potential, total organic carbon measurements and cryo-FEGSEM observations. We focus on two examples: the effect of the quantity of ammonium polyacrylate as dispersant and the effect of its chain length on alumina particles dispersion with a solid content of 32vol%. In the first example for some suspensions we measure values of viscosity or zeta potential too similar to discriminate the best state of dispersion. To overcome this problem we directly observe the suspensions with Cryo-FEGSEM. We take advantage of the recent developments of the technique, which provide now extremely high cooling rates and ensure that the freezing step does not induce observations artifacts related to the formation of ice. This technique provides an accurate vision of particles dispersion, agglomeration in ceramic suspensions and it is possible to visualize the excess of dispersant. In the second example the longer dispersant appears to be the more effective to obtain the best state of dispersion. Through both examples, we demonstrate that to have the best interpretation of results it is useful to combine direct observations by Cryo-FEGSEM and the usual properties measurements.

[*] e-mail: audrey.lasalle@saint-gobain.com

# Introduction

Cryo-SEM is an analytical technique which is widely used by biologists or chemists [1-4], or by the food industry [5-8], to observe hydrated samples without destroying their initial structure in the presence of water. In materials science, it has already allowed investigating suspensions with particles such as silica or latex [9-11], the hydration state advancement of cement [12], or the structure of gels [13]. The knowledge of the dispersion state of suspension is indeed a key parameter for many ceramic processing routes such as slip coagulation casting, fabrication of homogeneous coatings or spray drying. In the case of alumina particles the dispersion was widely studied by many authors[14-20].

Observations by cryo-SEM comprise a freezing step, used to fix the structure of the sample. During this step, the sample could be damaged by the ice crystals growing and thus the segregation of species in suspension, a phenomena well known in freeze-casting investigations and applications[21]. To overcome this problem vitrified water must be produced. Among the available techniques to obtain vitrified water, the high-pressure freezing (HPF) is the only method producing vitrified water on a depth of 200µm, by using a combination of very high cooling rate and high pressure, without using cryoprotectant. This technique has been first suggested by Moor & Riehle[22] in 1968 and then discussed and widely explained[23-25].

The creation of HPF system therefore greatly facilitate the sample preparation and ensure that the subsequent observations are free of freezing-induced artifacts. It is used to cryo-fix the sample with a thickness between 100-500µm without pre-treatment before Cryo-SEM observation.[9,11,26,27]

Measurements of zeta potential and viscosity, titration of adsorbed dispersants or sedimentation tests are common methods used to characterize the behavior of ceramic particles in suspensions[14,28,29]. By using AFM colloidal force measurements[15] or by calculating the diffuse layer and polymer adsorption layer thickness[20], it is possible to estimate steric, electrostatic or mixed repulsive forces responsible for dispersion phenomena. The main drawback of all these techniques is that they provide only indirect information on the dispersion state. In some cases, ceramic suspensions exhibit similar values of zeta potential and viscosity. It then becomes difficult to differentiate them, an issue rarely mentioned[30]. Direct observations usually consists in optical microscopy [30] or TEM observations [31,32] which requires low solid loadings or gel-casted suspensions.

Moreover with optical microscopy, images can be obtained only at low magnifications so that particles are not individually separated.

In this paper we use Cryo-FEGSEM to get a direct 2D image of the local organization of submicrometer alumina particles in aqueous suspensions with dispersants at a spatial resolution of a few nanometers. The range of optimum quantities of dispersants is preliminary determined by sedimentation tests. Then, Cryo-FEGSEM gives us an access to the particles morphology, the dispersion state, the aggregates formation, as well as the presence of an excess of organic additives. Titrations of the excess of dispersant are performed to validate Cryo-FEGSEM observations. This paper does not investigate the dispersion mechanisms of alumina particles but we use two classical examples to demonstrate here that cryo-images are beneficial to really assess the suspensions quality in addition to classical zeta potential, rheological behavior and pH measurements.

## I. Experimental

Experiments are carried out with a α-Alumina (Ceralox SPA 0.5, Sasol, Tucson, AZ, USA), dispersed in distilled water containing surfactants (table 1). This powder has a specific area (BET, Nova 2000, Quantachrom, FL, USA) of 7.4 $m^2g^{-1}$ and an average particles size of 0.4 µm (Sedigraph 5100, Micromeretics, GA, USA). Two dispersants are tested to generate negatively charged particles: Darvan CN and Darvan 821A (Vanderbilt, Norwalk, CT, USA) respectively referred as $D[NH_4^+]$ and $d[NH_4^+]$ in this paper. Darvan CN has a longer chain length than Darvan 821A, justifying the notations D[ ] and d[ ]. The counter ion, $[NH_4^+]$, is the same for both dispersants. Darvan CN and Darvan 821A are ammonium salts of polymethacrylic acid and polyacrylic acid with a molar weight of 10000-16000 g/mol and 3500 g/mol respectively (table 2). All prepared suspensions contain 32 vol% of alumina particles. For the two kinds of dispersants used, the ranges of the amounts of dispersants are chosen on the basis of preliminary sedimentation tests, not described here.

Dispersants are mixed with distilled water under magnetic stirring for 15 minutes. The powder is then added and the suspension ball-milled for 41 hours. The concentration of dispersant is given in milligrams of dispersant per $m^2$ of particles surface ($mg.m^{-2}$). A suspension dispersed for instance with 0.26 $mg.m^{-2}$ of the dispersant $D[NH_4^+]$ with respect to the surface area of alumina powder will be referred as 0.26 $mg.m^{-2}\_D[NH_4^+]$.

Viscosity measurements are performed in a concentric cylinder system (Bohlin viscosimeter, Malvern, Worcestershire, UK). The suspension is pre-sheared for 30 s followed by 30 s at rest. Viscosity is measured at a constant gradient of 50 s$^{-1}$. Zeta potential and pH are measured on the same suspensions as the ones used for viscosity measurements.

Zetaprobe (Colloidal Dynamics, North Attleboro, MA, USA) allows calculating the zeta potential on high solid loading suspensions. Traditional zeta potential techniques require sample dilution which likely affects the dispersion state of the suspension. The equipment used here is based on the electro-acoustic spectral analysis and can be used with suspensions up to 60 vol% solid loading. An electric field, applied between two probes in the suspension, creates sound waves inducing particles movement. Both electrical field and particles velocity are a sinusoidal signal. Zetaprobe measures the dynamic mobility of particles from the phase lag between both signals. Particles mobility depends on zeta potential and on an inertia factor, typical of the particles size. It is not necessary to know the particle size because the apparatus measures the inertia factor directly. The zeta potential is then directly calculated by the software. Both probes are dipped into 30 ml of de-aired alumina suspension poured into a Teflon container. Five consecutive measurements without rest are performed in order to avoid sedimentation effects.

Excess of dispersant is obtained by measuring the TOC, Total Organic Carbon (VCSN, Shimadzu, France) in the supernatant of a centrifuged suspension. Measurements consist in a catalytic combustion of the sample at high temperature. Carbon dioxide produced during the reaction is detected by the characteristic absorption of $CO_2$ in the infra red. The total carbon is obtained after the complete combustion of organic and inorganic compounds at 680°C, catalyzed by cobalt and platinum. The inorganic carbon is obtained after acidification and de-airing and the Total Inorganic Carbon results of the difference between the total and inorganic carbon. By knowing the TOC we have for a given quantity of dispersant, we can calculate the quantity of dispersant adsorbed on the particles.

To observe alumina particles into the suspension, a Jeol 7400 FEGSEM (Jeol, Tokyo, Japan) equipped with a Gatan Alto 2500 high resolution cryo stage (Gatan, Pleasanton, CA, USA) is used. It allows observing frozen suspensions. The freezing is performed using the Leica EMPact high pressure freezing device (Leica, Vienna, Austria) which provides a freezing pressure of 2000 bars with liquid nitrogen jets. The very fast cooling rate of the order of $10^4 K.s^{-1}$ and the increase of water viscosity combined with a sample thickness of 200µm leads to the formation of vitrified water and thus prevents from segregation phenomena. The high pressure of 2000 bars is responsible for the formation of vitrified water as shown in figure 5 of reference [33] and in figure 4 of reference [34]. It prevents the volume expansion of water due to liquid to solid transformation, limits ice nucleation rates and therefore the ice crystal formation. By using high pressure freezing system vitrified water or nanocrystals ice are produced[35,36]. If nanocrystals ice may appear, their size will not affect the spatial arrangement of the alumina particles because their size is larger by two orders of magnitude, and because the length of interactions of particles is larger than few nanometers because of the steric effect of dispersant. It is well-known and demonstrated that during the solidification of colloidal suspensions, there is a critical velocity of the solidification interface above which particles are entrapped by the moving interface, without moving or redistributing[37-41]. This critical velocity depends to some extent of the particle size. The cooling conditions used here, combined with the high pressure, leads to interfacial velocity larger than the critical velocity by several orders of magnitude. The spatial arrangement of particles is then kept as it was in the liquid. Three situations can occur: (i) the initial water is entirely vitrified (ii) most of the water is vitrified, and ice nanocrystals are formed (iii) no vitrified water is formed; only ice crystals are present. Based on the experimental conditions and samples characteristics used here, we believe we are in either situation (i) or (ii). The occurrence of such situations has been mostly investigated in the case of the preparation of biological samples. For instance, Studer & al[35] have observed the state of water (vitrified or crystallized) by electron diffraction after high pressure freezing of bovine articular cartilage. They calculated than a tissue sample with a thickness of 200µm freezing at $10^4 K.s^{-1}$ under 2100 bars has a cooling rate inside the sample close to 5000 $K.s^{-1}$. To be vitrified this cooling rate must be between $10^3$ and $10^5$ $K.s^{-1}$. We can thus safely assume than by using a cooling rate of $10^4 K.s^{-1}$ under 2100 bars to cryo-fix alumina suspensions on a depth of 200µm, and considering that thermal conductivity of alumina is about 30 times that of water, the cooling rate will be sufficient to produced only vitrified water.

To prepare the sample a droplet of the alumina is pipetted into a specimen carrier (diameter 1.2mm, depth 200µm) which is placed into a pod and tightens securely with the supplied torque wrench. The assemblage is loaded into the EM pact high-pressure machine and subjected to program cooling by liquid nitrogen jets so that the suspension is vitrified at $10^4 K.s^{-1}$, under a 200 MPa pressure. These parameters are checked on the control-screen of the EM pact high-pressure machine. The extreme cooling rate from ambient temperature to liquid nitrogen temperature under high pressure leads to the formation of vitrified water. After cryo-fixation the sample is stored into a liquid nitrogen bath at -196°C to keep sample in its amorphous state. Cryo-fixed sample is placed on the cold stage located outside of the SEM, fractured with a sharp knife and the surface is sublimed at -95°C during 15 minutes under secondary vacuum to increase topological contrast. Finally the sample is transferred in the SEM where the cold stage, maintained at -100°C, keeps samples frozen during observation, avoiding as much as possible the nucleation of ice crystals on sample surface and the sublimation which could damage the structure of the sample. Secondary electron images of the sample are acquired at a low accelerating voltage of 2keV.

## II. Results & discussion

Two examples are given here to illustrate the interest of Cryo-FEGSEM in addition to other analytical techniques to accurately characterize the dispersion state of alumina suspensions: *(i)* effect of the quantity of a polyelectrolyte dispersant and *(ii)* effect of its chain length.

### (1) Effect of D[$NH_4^+$] dispersant quantity

When dispersed in distilled water, with no polyelectrolyte addition, the natural pH of our alumina is around 8.7. When a quantity of 0.26 to 0.51 mg.m$^{-2}$ D[$NH_4^+$] is added (optimum range defined upon preliminary sedimentation tests), the natural pH increases up to 9.1 and 9.3, respectively. Such a variation of natural pH between minimum and maximum quantities of added dispersant is not really significant. However, we know that, in this pH range, the alumina surface is positively charged, whereas the D[$NH_4^+$] polyelectrolyte is negatively charged with a train and loops configuration. The major part of the chain of the dispersant thus extend into the liquid but a part of it is also linked to the particles surface through the COO$^-$ functional groups [14]. The non-adsorbed COO$^-$ groups lead to pending negative charges around the particles resulting in high negative values of zeta potential.

In the range 0.26-0.51 mg.m$^{-2}$ of D[$NH_4^+$], all alumina suspensions have absolute values of zeta potential over 68 mV (figure 1a), which is usually considered as characteristic of a good stability [29]. As expected, their viscosities are very low, ten times lower than with a lower quantity of dispersant of 0.13 mg.m$^{-2}$ (figure 1b*)*. These results are in good agreement with those of Wei et al for similar alumina powder and dispersant [28]. We observe an increase of viscosity with 0.31mg.m$^2$ of D[NH4+] followed by a decrease and a new increase. This phenomena has already been reported by Davies &al[14], although with a greater amplitude. They propose a change of configuration of the dispersant on particle surface due to the counter–ion concentration. In the specific cases of 0.26 and 0.39 mg.m$^{-2}$ D[$NH_4^+$] suspensions, the zeta potential and viscosity values are almost similar, less than 5mV and 5mPa.s$^{-1}$ between values, so that it is impossible to determine the best dispersion state. Takahashi *et al* faced the same problem on alumina suspensions with lower solid loadings (*i.e.* 10 wt%) and they proposed to gel-cast the suspension, dry it, slice the sample and finally observe it by optical microscopy or SEM [30]. In our case, we use the Cryo-FEGSEM to determine which formulation brings to the best dispersion state. Cryo-FEGSEM images show that, without dispersant (figure 2a, e) particles are strongly agglomerated. This phenomenon is observable by the formation of large empty spaces, indicated by black and white arrows on each image, between agglomerates (surrounded by a dashed line). With 0.26 mg.m$^{-2}$ of D[NH$_4^+$], particles are well separated, with no visible agglomerates (figure 2b), no large empty spaces appear between particles (figure 2f), so that the solid loading seems to be higher than 32vol%. This can be explained by the depth of field of SEM: several levels of particles repartition could be observed. When the solid content of the suspension increases (figure 3) the number of observable layers decreases. When the solid content increases up to the maximal packing of alumina particles (figure 3 c) only the upper layer of particle is visible.

When more dispersant is added, such agglomerates can however be clearly identified from 0.39 mg.m$^{-2}$ (figure 2c). Empty spaces (pointed out by black and white arrows) appear between particles, such a trend being even slightly reinforced for 0.51 mg.m$^{-2}$ of D[NH$_4^+$]. It has to be noticed that filaments, pointed out by a white arrow on figure 2h and also observable in figure 2g, seem to cover some particles or link them for 0.39 mg.m$^{-2}$ and 0.51 mg.m$^{-2}$ of D[NH$_4^+$]. A similar observation was made by Wyss *et al* on a silica particles suspension containing a too high urea concentration [9]. The filaments are the result of a reaction which takes place during the sample sublimation before observation when there is an excess of organic compounds in the aqueous media. On figure 2g and 2h particles seem to be less numerous than in figure 2f. It could be due to a slight sedimentation effect caused by the excess of dispersant.

Hence, our best dispersion state is obtained for 0.26 mg.m$^{-2}$ of D[NH$_4^+$]. Above 0.26 mg.m$^{-2}$, some of the added dispersant is not adsorbed anymore onto the alumina surface as confirmed by TOC measurements (figure 4). Actually, the alumina surface is completely saturated with an added amount of D[NH$_4^+$] dispersant of about 0.39 mg.m$^{-2}$, so that the added dispersant beyond this value remains in excess in the liquid (*see* plateau on figure 4). The consecutive increase of ionic strength logically leads to a destabilization which is correlated to a measurable viscosity increase only from 0.51 mg.m$^{-2}$ of D[NH$_4^+$] (figure 1 b), whereas it is observed as early as 0.39 mg.m$^{-2}$ by cryo-FEGSEM. The viscosity measurements are normally very sensitive to the dispersion state but only for high solid loadings (above 50 vol%) which is not the case here. Finally, the registered slight but significant decrease of absolute values of zeta potential between the 0.39 mg.m$^{-2}$ and 0.51 mg.m$^{-2}$-D[NH$_4^+$] suspensions (figure 1a) may be attributed to a modification of the double layer structure around solid particles. Indeed, a strong increase of the ionic strength due to non-adsorbed polyelectrolyte species usually tends to compress the double layer.

## (2) Effect of dispersant chain length

The $D[NH_4^+]$ dispersant has a chain length three times longer than that of $d[NH4^+]$ with exactly the same counter ions and functional groups sequence attached to the hydrocarbon backbone. This is the reason why it remains significant to compare, like we do, the efficiency of those two dispersants on the basis of their mass concentrations with respect to the powder surface. Our data in mg.m$^{-2}$ may be anyway easily transformed into moles.m$^{-2}$ knowing that there is roughly a ratio of 3 in term of number of moles between the two dispersants for a given mass.

The figure 4 shows that the alumina surface is saturated with $D[NH_4^+]$ starting from a mass concentration of 0.39 mg.m$^{-2}$, while no saturation (*i.e.* no plateau) is observed with $d[NH_4^+]$ within the tested mass concentration range. At the same time, the absolute values of zeta potential are higher with $D[NH_4^+]$ than with $d[NH_4^+]$, especially at low mass concentrations of dispersants (figure 5a). We can reasonably conclude that $D[NH_4^+]$ is a more efficient dispersant than $d[NH_4^+]$. This could be explained by the large size of the $D[NH_4^+]$ molecule in combination with its specific molecular conformation as train and loops. Both could promote a good coverage of the alumina surface resulting in a high packing density of COO$^-$ negative pending groups responsible for high absolute values of zeta potential.

On the contrary, in the case of $d[NH_4^+]$, for a given added mass of dispersant, a higher fraction of it is adsorbed onto the surface of alumina than in the case of $D[NH_4^+]$ (figure 4), but the resulting absolute values of zeta potential are lower (figure 5a). It may indicate that due to the $d[NH_4^+]$ molecular conformation, a higher ratio of available COO$^-$ functional groups takes part to the surface bonding, leaving a lower amount of pending COO$^-$. In parallel, it increases the bonding strength of elementary $d[NH_4^+]$ molecules, thus increasing their stability. The fact that the $d[NH_4^+]$ grafting lets less numerous pending charges also explains why, for a given added mass of dispersant, a larger amount of it can be adsorbed. Indeed, pending negative charges may counteract the bonding formation through repulsion forces.

From the range of quantity of dispersant tested, we choose to compare suspensions with 0.26 and 0.51 mg.m² of dispersant because the first corresponds to the best state of dispersion when D[NH4+] and the second because zeta potential and viscosity values are almost equal. Anyway, even for 0.26 mg.m$^{-2}$ of d[NH$_4^+$], the absolute value of zeta potential is high enough to promote in principle a good dispersion, which is exactly what we observe by cryo-FEGSEM (figure 6c, d). When the quantity of added d[NH$_4^+$] increases up to 0.51 mg.m$^{-2}$, the absolute value of zeta potential also increases and the good dispersion state is kept (figure 6g, h). Contrary to what is observed with 0.51 mg.m$^{-2}$ of D[NH$_4^+$], no agglomeration occurs. This might be due to the fact that with d[NH$_4^+$], a larger amount of added dispersant is grafted so that less numerous electrically charged molecules remain in excess in the liquid (no filaments observed on figure 6h), limiting the destabilization due to a potential rise of the ionic strength.

Another interesting point is that the viscosity of suspensions dispersed with d[NH4+] is higher or equal than with D[NH4+] (figure 5b) whereas cryo-FEGSEM reveals a better dispersion state than suspensions dispersed with more than 0,26mg.m-2 of D[NH4+] (figure 6 d, h and f). With 0.26mg.m² of d[NH4+] the suspension is more viscous but shows a similar state of dispersion than the suspension with the same quantity of D[NH4+] (figure 6 b and d). More large particles are visible on the first layer of particles with D[NH4+] than with d[NH4+]). We know that the interaction potentials are basically a combination of electrostatic and steric ones. In case of d[NH4+] the chain length is three times smaller than D[NH4+], thus these the chain length and the quantity maybe too short and low enough to maintain the bigger particles on surface without impacting dramatically the state of dispersion. Of course, when the quantity of d[NH$_4^+$] increases, the absolute value of zeta potential increases together with the interaction forces leading to a lowering of the viscosity which becomes similar for the two types of dispersants at 0.51 mg.m$^{-2}$ (figure 5b).

We demonstrate here again that, for suspensions with intermediate solid loadings, cryo-FEGSEM provide several additional information's on the dispersion state.

## III. Conclusion

Based on the investigation of the stability of alumina suspensions by different techniques, we show here that cryo-FEGSEM is a good tool to obtain additional information on the state of dispersion of alumina suspensions, necessary to conclude and get an accurate vision of the local state of dispersion of particles. It is particularly true for suspensions with low or intermediate solid loadings for which the viscosity is at the same time too low and too sensitive to the local arrangement of particles to be a discriminating parameter. Interestingly Cryo-FEGSEM provides useful images which can be analyzed:
- to discriminate the suspensions for which zeta potential and viscosity values are very similar

- to observe the formation of agglomerates
- to observe the excess of dispersant in suspension characterized by the formation of filaments between particles.

To have a better understanding of efficient dispersion mechanisms, and to conclude on the real state of a particulate dispersion, the best analytical procedure will consist in combining zeta potential and viscosity measurements with Cryo-FEGSEM observations and titration of adsorbed quantity of dispersants. Cryo-FEGSEM does not replace classical analytical techniques but provides unique complementary information.

| Crystal phase | α-Al$_2$O$_3$ |
|---|---|
| Average particles size d$_{50}$ (μm) | 0.4 |
| BET surface area (m²·g$^{-1}$) | 7.4 |
| Particle density (g·cm$^{-3}$)* | 3.96 |
| Zeta potential without dispersant (mV) | 20 |
| Natural pH | 8.7 |
| IEP | 9.85 |

**Table 1.** Alumina properties measured or provided by the manufacturer*.

| Dispersant name | Notation | Chemical formula | Molar weight (g·mol$^{-1}$) |
|---|---|---|---|
| Darvan CN | D[NH$_4^+$] | [C$_3$H$_5$COO$^-$]$_n$[NH$_4^+$]$_n$ | 10000 -16000 |
| Darvan 821A | d[NH$_4^+$] | [C$_2$H$_4$COO$^-$]$_n$[NH$_4$]$^+_n$ | 3500 |

**Table 2.** Dispersant characteristics.

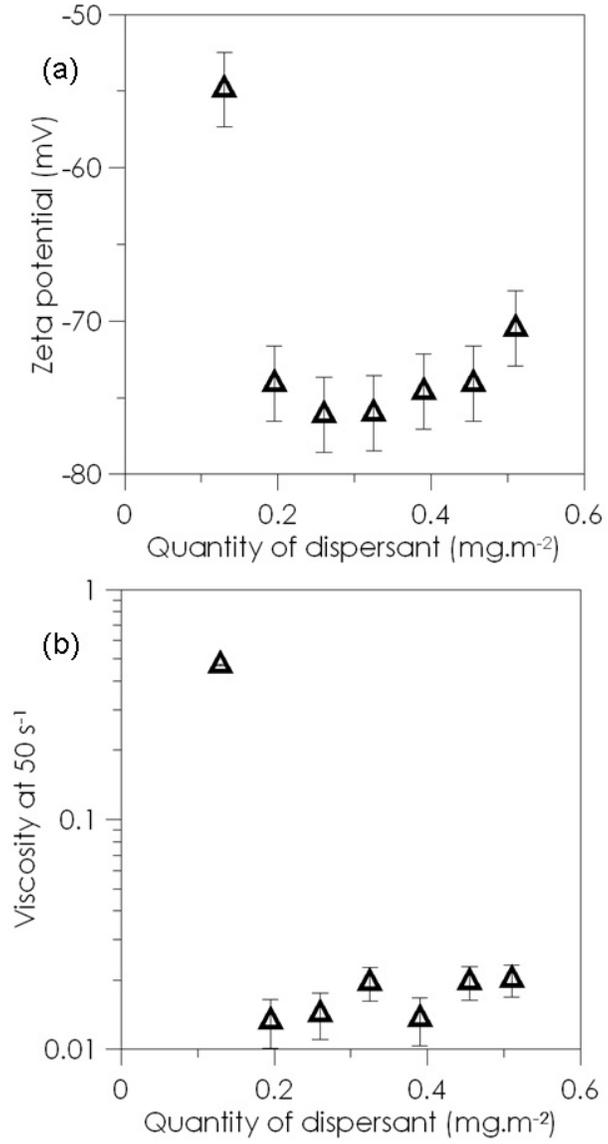

**Figure 1.** Zeta potential (a) and viscosity values (b) of alumina suspensions dispersed with D[$NH_4^+$].

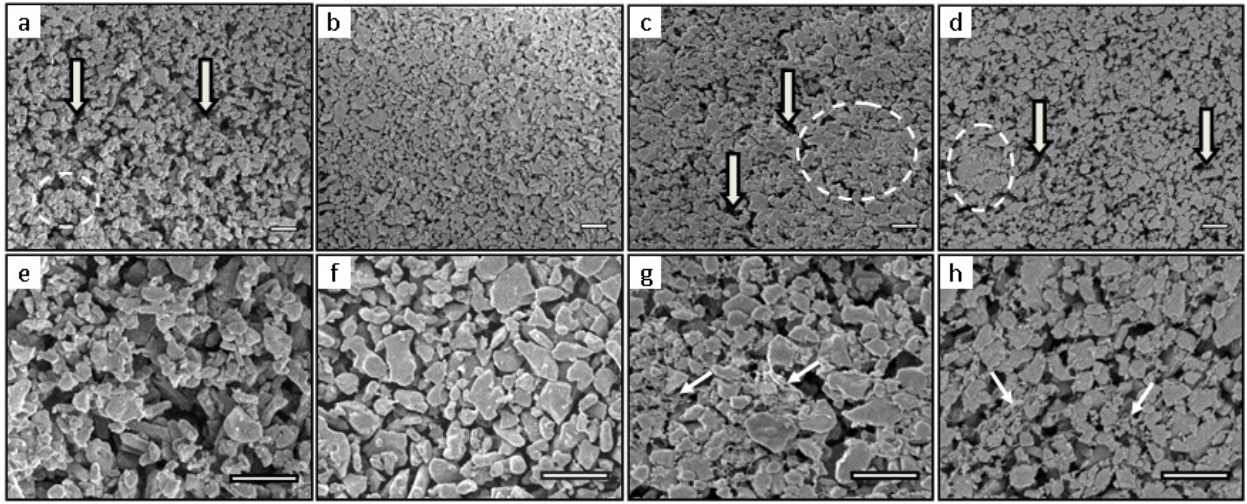

**Figure 2.** Cryo-FEGSEM pictures of suspensions without dispersant, with 0.26mg.m$^{-2}$, 0.39mg.m$^{-2}$ and 0.51mg.m$^{-2}$ of D[NH$_4^+$] at magnification x 10000 (a-d) and x 25000 (e-h). Cryo-FEGSEM pictures show a well dispersed state when 0.26mg.m$^{-2}$ of D[NH$_4^+$] is added (b,f) whereas an excess of dispersant, pointed out by white arrows, creates agglomerates (pointed out by black & white arrows) and spaces between particles (surrounded by a dashed line), appears when the amount of D[NH$_4^+$] increases up to 0.39 mg.m$^{-2}$ (c) and 0.51mg.m$^{-2}$ (d). Scale bar: 1 µm.

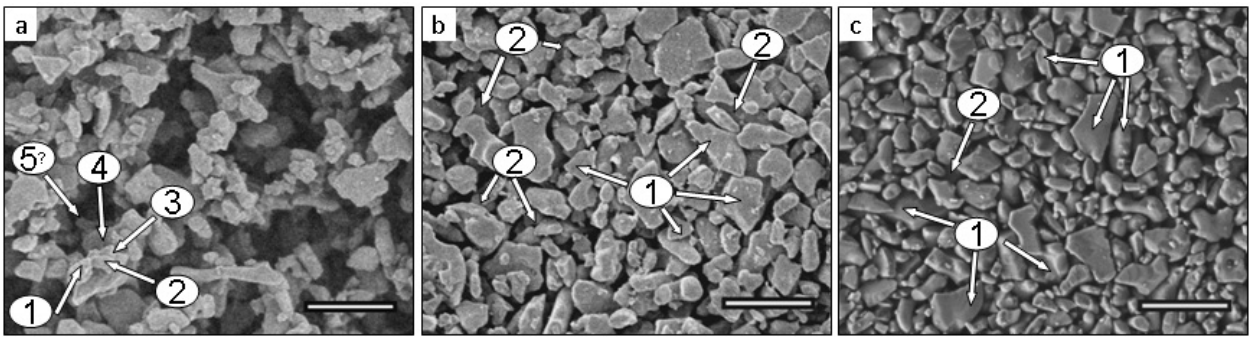

**Figure 3. Effect of** depth of field of Cryo-FEGSEM images according to the solid content, 11vol% (a), 32vol% (b), 56vol% (c). When the solid content increases up to the maximal packing of alumina particles, the number of observable particles layers (pointed out by black and white arrows and numbered from 1 to 5) decreases. Scale bar: 1 µm

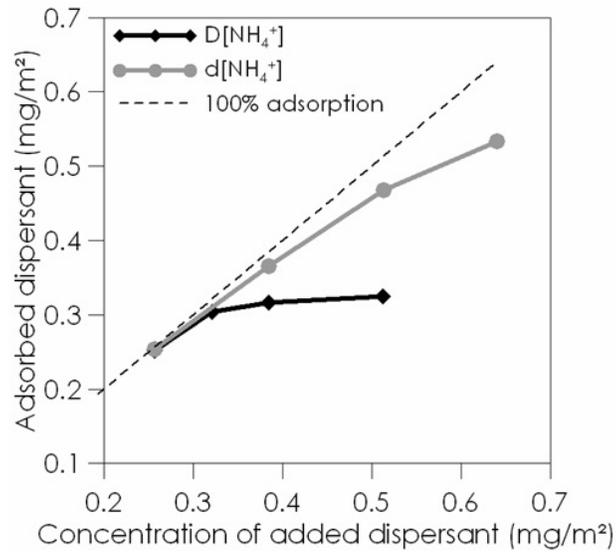

**Figure 4.** TOC measurements expressed as the variation of the quantity of adsorbed dispersant as a function of the initial added quantity. The case where all dispersant is adsorbed is represented by the dashed line.

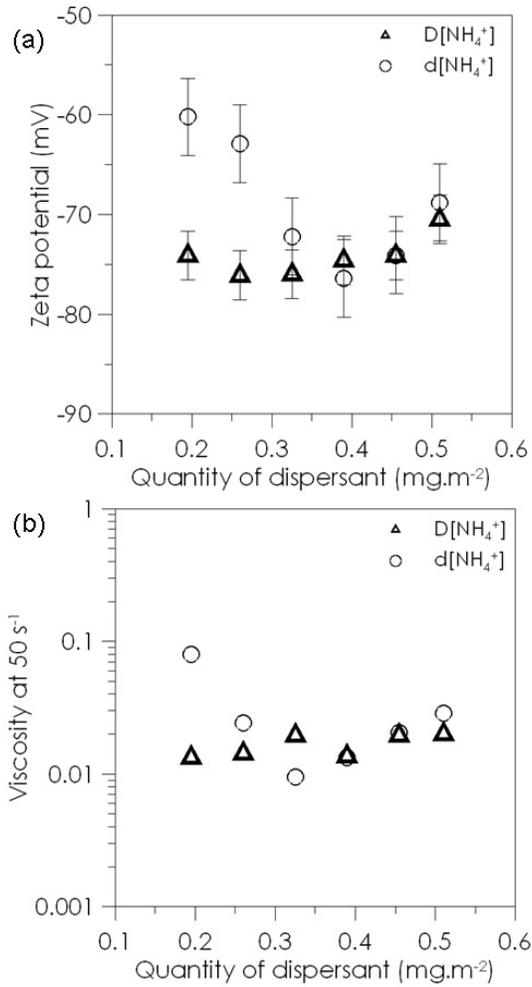

**Figure 5.** Effect of the chain length of dispersant on the zeta potential values (a) and viscosity measurements (b).

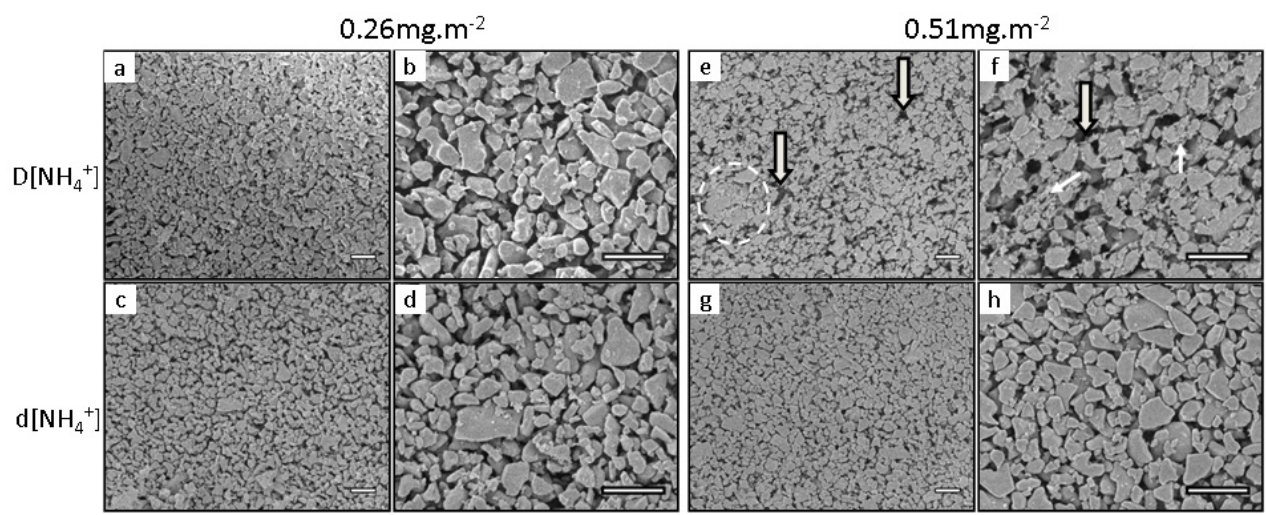

**Figure 6.** Cryo-FEGSEM pictures of alumina suspensions dispersed with 0.26mg.m$^{-2}$ of D[NH$_4^+$] (a-d), 0.51mg.m$^{-2}$ of d[NH$_4^+$] (e-h), magnification X10000 for (a, c, e, g) and x25000 for (b, d, f, h). Every suspension is well dispersed except 0.51mg.m$^{-2}$-D[NH$_4^+$] suspension where excess of dispersant pointed out by a white arrow creates agglomerates, surrounded by dashed line. Scale bar: 1 µm.